\documentclass[twocolumn,showpacs,showkeys,preprintnumbers,amsmath,amssymb]{revtex4}

\usepackage{nomencl}
\usepackage[latin1]{inputenc}
\usepackage[active]{srcltx}
\usepackage{setspace}
\usepackage{graphicx}
\usepackage{dcolumn}
\usepackage{bm}
\usepackage{epsfig}
\usepackage{epstopdf}
\usepackage{marvosym}
\usepackage{wasysym}

\makenomenclature
\begin{document}

\preprint{Preprint IST/DF 2015-M J Pinheiro}


\title[]{Effect of TTC on Satellite Orbital Mechanics}

\author{Mario J. Pinheiro}
\email{mpinheiro@tecnico.ulisboa.pt}

\affiliation{Department of Physics, Instituto Superior T\'{e}cnico - IST, Universidade de Lisboa - UL,
Av. Rovisco Pais, 1049-001 Lisboa, Portugal}

\homepage{http://mjpinheiro.weebly.com/}

\thanks{}

\date{\today}

\begin{abstract}
The modified dynamical equation of motions introduced in previous publication topological torsion current (TTC) [Mario J. Pinheiro (2013) 'A Variational Method in Out-of-Equilibrium Physical Systems', Scientific Reports {\bf 3}, Article number: 3454] predicts a so-far unforseen anomalous acceleration detected in spacecrafts during close planetary flybys in retrograde direction, and a null-effect when the spacecraft approach the planet in posigrade direction.
\end{abstract}



\maketitle

\section{Introduction}

We suggest a possible theoretical explanation of the physical process underlying the unexpected orbital-energy change observed during the close planetary flybys~\cite{Anderson 2008,Turyshev 2009} based on the topological torsion current (TTC) found in a previous work~\cite{Pinheiro_SR}. The theoretical framework is classical. If the proposed theory is just a prototype or a complete framework to describe the phenomena, it remains to be seen, after minute comparison with orbital data. The dynamics follow according the set of two first order differential equations (see Ref.~\cite{Pinheiro_SR}):
\begin{align}
\partial_{\mathbf{p}^{(\alpha)}} \bar{S} \geq 0 \label{g2b}\\
\nonumber \\
\partial_{\mathbf{r}^{(\alpha)}} \bar{S} = -\eta \partial_{\mathbf{r}^{(\alpha)}} U^{(\alpha)} - \eta m^{(\alpha)} \partial_t \mathbf{v}^{(\alpha)} \geq 0  \label{g2a}.
\end{align}
Here, $\eta \equiv 1/T$ is the inverse of the ``temperature" (not being used so far), and we employed condensed notation: $\partial_{\mathbf{p}^{(\alpha)}} \equiv \partial/\ \partial_{\mathbf{p}^{(\alpha)}}$. Then, we obtain a general equation of dynamics for gravitational systems~\cite{Pinheiro_SR}:
\begin{equation}\label{eq7b}
m \frac{d \mathbf{v}}{d t} = - \pmb{\nabla} \phi + m [\mathbf{A}  \times \pmb{\omega}].
\end{equation}
The first term in the rhs is the usual gradient of the gravitational potential at zero-order of its expansion in spherical harmonics
\begin{equation}\label{eq7b1}
\phi = - \frac{GM_{\oplus}m}{r},
\end{equation}
and the last term of Eq.~\ref{eq7b} represents the topological torsion current~\cite{Pinheiro_SR} (TTC), which includes the gravitational vector potential $\mathbf{A}$, given in the form of a retarded potential
\begin{equation}\label{eq8}
\mathbf{A}=\frac{G}{c^2}\frac{M\mathbf{v}_{SP}}{|\mathbf{r}-\mathbf{r}'|\left( 1-\frac{(\mathbf{v}_{SP} \cdot \mathbf{n}')}{c} \right)}.
\end{equation}
We may notice that the topological torsion current is the outcome of a balance between: i) a minimum of mechanical energy and, ii) a maximum of entropy. The fundamental equation of dynamics written in the form of Eq.~\ref{eq7b} shows that TTC only appears in physical systems that are out-of-equilibrium (i.e., accelerated) and is a consequence of a balance between energy and entropy, not just the criterium of minimization of energy, as usually imposed, and therefore may be able to contains new physics.

We hereby introduce some useful nomenclature: $m$ = mass of spacecraft; $M_{\oplus}$ = Earth's mass ($\earth$); $G$ = Gravitational constant; $c$ = speed of light; $\omega_{\oplus}$ = Earth's angular velocity of rotation; $R_{\oplus}$ = Earth's radius; $\mathbf{v}_{SP}$ = velocity of the sun relative to the planet; $V_P$ = velocity of the planet relative to the Sun; $v_{\infty}$ = excess hyperbolic speed of the spacecraft relative to the planet; ($v_{xo}$, $v_{yo}$, $v_{zo}$) = components of the spacecraft velocity in the referential ($x_o$, $y_o$, $z_o$); $\mathbf{h}$ = angular momentum normal to the plane of the orbit.

To simplify the complex mathematical problem, we assume the geometry of Fig.~\ref{Geo}, the cylindrical coordinate frame. The second term in the right hand side of Eq.~\ref{eq7b} in the chosen geometry ($\pmb{\omega}=\omega_{\oplus} \mathbf{K}$), gives
\begin{equation}\label{eq9}
[\mathbf{A} \times \pmb{\omega}] = A_{xo} \omega_{\oplus} [\mathbf{x}_o \times \mathbf{K}]+A_{yo} \omega_{\oplus}[\mathbf{y}_o \times \mathbf{K}]+A_{zo}[\mathbf{z}_o \times \mathbf{K}].
\end{equation}
Then, we obtain
\begin{equation}\label{eq10}
[\mathbf{A} \times \pmb{\omega}]=\omega_{\oplus} \sin I(A_{xo}-A_{zo}) \mathbf{y}_o.
\end{equation}
Now, we develop the retarded gravitational vector potential to the first term in a Taylor series, since, as we will see, it is the non-instantaneous character of the gravitational interaction together with TTC that contributes to the flyby anomaly:
\begin{equation}\label{eq11}
\mathbf{A} \approx \frac{GM_{\oplus}}{c^2}\frac{\mathbf{v}}{r}+ \frac{GM_{\oplus}}{c^2}\frac{\mathbf{v}}{r}\left( \frac{\mathbf{v} \cdot \mathbf{n}^{'}}{c} \right),
\end{equation}
or considering that $\mathbf{n}^{'} \equiv \mathbf{u}_r$, $(\mathbf{v} \cdot \mathbf{n}^{'})=v_r$ is the relative speed of the spacecraft to Earth, and assuming that $v_r \simeq V_P \simeq v_{apx}$, then Eq.~\ref{eq11} takes the form
\begin{equation}\label{eq12}
\mathbf{A} \approx \frac{GM_{\oplus}}{c^2 r}(v_{xo} + \frac{v_{xo}v_r}{c})\mathbf{x}_o + \frac{GM_{\oplus}}{c^2r}(v_y+\frac{v_{yo}v_r}{c})\mathbf{y}_o.
\end{equation}

Therefore, it is obtained the approximate set of differential equations (note that $A_{zo}=0$) governing the spacecraft during the flyby manoeuver:

\begin{figure}
  \centering
  \includegraphics[width=3.0 in]{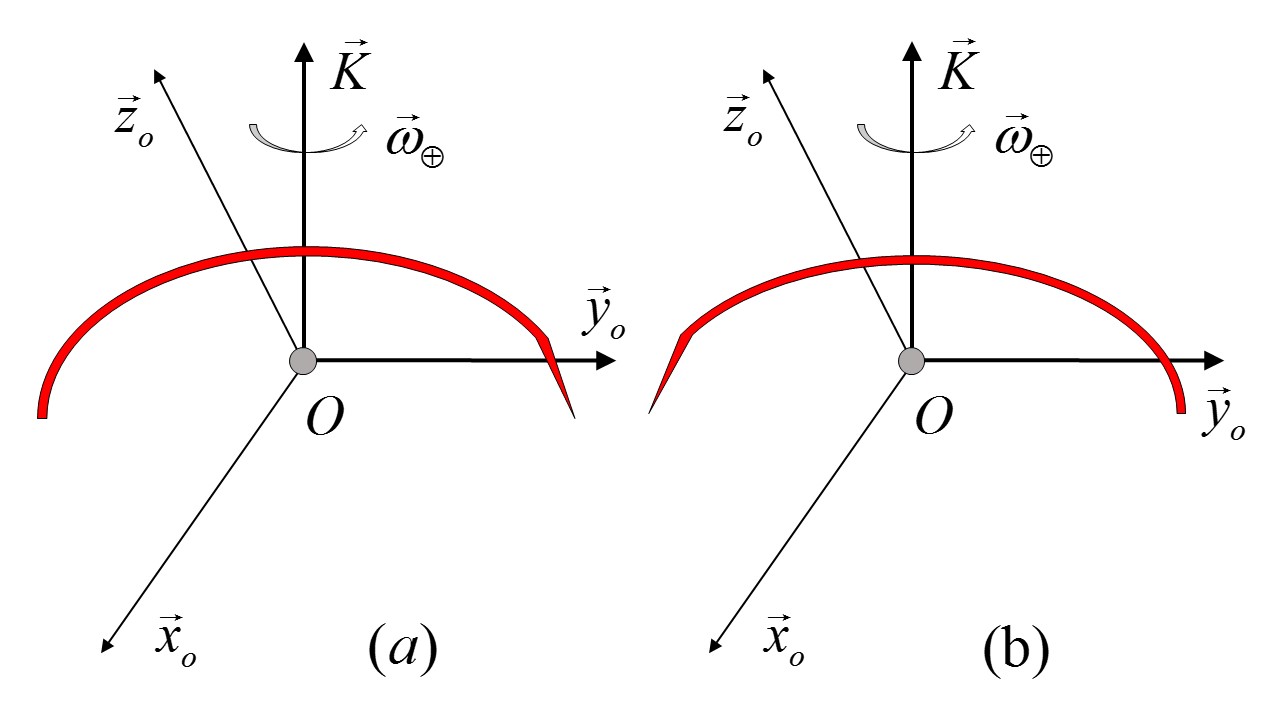}\\
  \caption{Cylindrical coordinate frame.}\label{Geo}
\end{figure}

\begin{eqnarray}\label{set}
  m\frac{dv_{xo}}{dt} &=& -\frac{GM_{\oplus}m}{r^2}(\mathbf{u}_r \cdot \mathbf{u}_{xo}) \\ \label{eq01}
  m\frac{dv_{yo}}{dt} &=& -\frac{GM_{\oplus}m}{r^2}(\mathbf{u}_r \cdot \mathbf{u}_{yo}) + m\omega_{\oplus} A_{xo} \sin I \\ \label{eom2}
  m\frac{dv_{zo}}{dt} &=& -\frac{GM_{\oplus}m}{r^2}(\mathbf{u}_r \cdot \mathbf{u}_{zo}) \label{eq03}
\end{eqnarray}

The approach velocity vector $\mathbf{v}_{ap}$ is expressed in the approach plane $(\mathbf{i},\mathbf{j},\mathbf{h})$ as follows (the unit vector $\mathbf{i}$ points along the planet direction of motion):
\begin{equation}\label{eq15de}
\mathbf{v}_{ap}=v_{apx} \mathbf{i} + v_{apy} \mathbf{j} + v_{apz} \mathbf{h},
\end{equation}
and the general representation of the spacecraft velocity vector relative to Earth in the direct orthogonal frame is given by
\begin{equation}\label{eq15df}
\begin{array}{ll}
  v_{apx}&=V_P + v_{\infty} \cos (\omega \mp \theta) \\
  v_{apy}&=v_{\infty} \sin(\omega \mp \theta) \\
  v_{apz}&=0.
\end{array}
\end{equation}

In consequence, according to Eq.~\ref{eq11}, we have
\begin{equation}\label{eq13}
A_{zo}=0,
\end{equation}
and the remaining terms are the following
\begin{equation}\label{eq14}
A_{xo}\approx \frac{GM_{\oplus}}{c^2r}v_{xo} + \frac{GM_{\oplus}}{c^2r}\frac{v_{xo} v_r}{c}.
\end{equation}
Now, to compactify the notation, we write $u\equiv \frac{GM_{\oplus}}{c^2}$ and $\theta^{'} \equiv \omega \mp \theta$. Therefore, Eq. 11 transforms into

\begin{equation}\label{eq16}
  m\frac{dv_{yo}}{dt} = m \omega_{\oplus} \frac{u}{r} v_{xo} \sin I + 2m\omega_{\oplus}\frac{u}{2r}\frac{v_{xo} v_{\infty}}{c}\cos \theta^{'} \sin I
\end{equation}
Now, we transform from the cylindrical coordinate frame to Hill coordinate frame, more appropriate to describe the orbital motion
\begin{eqnarray}
  \mathbf{x}_o &=& \mathbf{u}_r \\
  \mathbf{y}_o &=& \mathbf{u}_{\theta} \\
  \mathbf{z}_o &=& \mathbf{h}
\end{eqnarray}

In the Hill coordinate frame, the set of Eqs.~\ref{set}-12 can now be written in the more useful form
\begin{eqnarray}
 m\frac{dv_r}{dt} &=& -(\nabla \phi)_r \\ \label{max3}
  m\frac{dv_{\theta}}{dt} &=& m\omega_{\oplus}\frac{u}{r}v_r\sin I+2m\omega_{\oplus}\frac{u}{r}\frac{v_r^2}{c}\cos \theta^{'}\sin I \\
m\frac{dv_z}{dt} &=& 0. \label{max5}
\end{eqnarray}

\subsection{Conditions for the occurrence of the flyby anomaly}

In a previous work~\cite{Pinheiro_PLA} we discuss the role of TTC, but now, we specialize that framework to give an explanation to the Juno flyby that registered unexpectedly no anomaly~\cite{Hori}.

As seen before, the equation to solve is
\begin{equation}\label{eq34}
\mathbf{\ddot{r}}=-\pmb{\nabla} \phi + \mathbf{f}_{TTC}.
\end{equation}
We will use the cylindrical coordinate frame $\mathcal{F}_0$ (see Ref.~\cite{Ruiter}) with basis vectors $(\mathbf{x}_o, \mathbf{y}_o, \mathbf{z}_o)$. In this coordinate frame, the position and velocity are given by
\begin{eqnarray}
  \mathbf{r} &=& r\mathbf{x}_o \\
  \mathbf{v} &=& \dot{r}\mathbf{x}_o+r \dot{\theta}\mathbf{y}_o
\end{eqnarray}
and the angular momentum is given by
\begin{equation}\label{eq35}
\mathbf{h}=h\mathbf{z}_o=r^2 \dot{\theta}\mathbf{z}_o.
\end{equation}
The perturbing force in $\mathcal{F}_o$ coordinates is given by
\begin{equation}\label{eq36}
\mathbf{f}_{TTC}=f_r \mathbf{x}_o+f_{\theta}\mathbf{y}_o+f_z\mathbf{z}_o.
\end{equation}

\subsubsection{Spacecraft in posigrade direction}

When the spacecraft enters in flyby maneuvering in posigrade direction, the initial conditions for the velocity components are (changing the variable to $\theta'\equiv \omega \mp \theta$ for compactness of notation):
\begin{eqnarray}
  v_{apx} &=& V_P - v_{\infty} \cos (\theta^{'}) \\
  v_{apy} &=& v_{\infty} \sin (\theta^{'}) \\
  v_{apz} &=& 0.
\end{eqnarray}
with
\begin{equation}\label{eq37}
\mathbf{n}^{'}=\sin(\theta^{'}) \mathbf{y}_o + \cos (\theta^{'}) \mathbf{x}_o
\end{equation}
and therefore, the term $(\mathbf{v} \cdot \mathbf{n}^{'})$ appearing in Eq.~\ref{eq11} transforms into (we solve now in the frame of the planet, hence $V_P=0$)
\begin{eqnarray}
  (\mathbf{v} \cdot \mathbf{n}^{'}) &=& v_{apx} \cos (\omega \mp \theta) + v_{apy} \sin(\theta^{'}) \\
   &=& v_{\infty} \cos^2 (\theta^{'}) - v_{\infty} \sin^2 (\theta^{'}) \\
    &=&  v_{\infty} \cos 2(\theta^{'}).
\end{eqnarray}

To solve the set of Eqs.~\ref{set}-~\ref{eq03} is not our intention due to its complexity, we just want to solve them in the zone where the flyby occurs, taking $R \approx R_{\oplus}$, and assuming certain approximations that simplify the mathematical problem at hand, namely that $v_{x_0} \approx 0$, and $\frac{v_{x_0} v_{\infty}}{v} \sim v_{\theta}$. Eq. 22 can be written in the form
\begin{equation}\label{eq38}
\frac{d v_{\theta}}{dt}=\frac{2 \omega_{\oplus} R_{\oplus}}{c}\frac{v_{x_0} c}{R_{\oplus}}\sin I + \frac{2 \omega_{\oplus} R_{\oplus}}{c}\frac{v_{x_0} v_{\infty}}{R_{\oplus}}\sin I
\end{equation}
Here, $K \equiv \frac{2 \omega_{\oplus}}{R_{\oplus}}$, or
\begin{equation}\label{4law}
\frac{d v_{\theta}}{dt}=K \frac{v_{x_0}c}{R_{\oplus}}+K\frac{v_{x_0}v_{\infty}}{R_{\oplus}}\sin I
\end{equation}
Eq.~\ref{4law} shows two forces acting together on the spacecraft. The first term is nearly zero, but not null, since we assume nearly circular trajectory. However, the second term of force actuates as a modulating force and contributes more significantly. Hence, assuming a nearly circular trajectory during the flyby, $dt=R_{\oplus}d\theta/v$. The equation of motion for the azimuthal velocity becomes
\begin{equation}\label{eq39}
\frac{dv_{\theta}}{d \theta}= \left( \frac{K c v_r}{v} \sin I + \frac{K v_{\infty} v_r}{c} \cos 2(\theta^{'}) \sin I \right) \cos (\theta^{'}).
\end{equation}
The mathematical solution of Eq.~\ref{eq39} is (and from now omitting the apostrophe for clarity)
\begin{equation}\label{eq40}
\begin{split}
v_{\theta}= \exp [c_2 \sin \theta + c_3 \sin (3 \theta)](c_1+\\
  c_4 \int_1^{\theta} \exp [-c_1 \sin (\xi)-
       c_5 \sin (3 \xi)] \cos \xi d \xi.
\end{split}
\end{equation}

\begin{figure}
 \centering
\includegraphics[width=3 in]{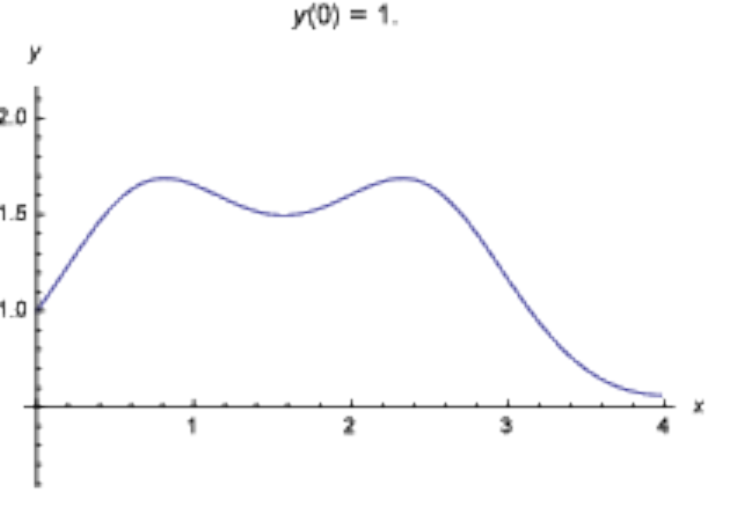}\\
\caption{Azimuthal velocity $y=v_{\theta}$ vs. $x=\theta$ (in rad units), from $0 \to \pi$, in the posigrade direction.}\label{Fig3}
\end{figure}

Fig.~\ref{Fig3} shows the velocity anomaly in the posigrade direction. It occurs an acceleration during half-way, with a bump at the middle of the orbit, followed by a deceleration, giving an null effect. Notice that the magnitude of the velocity anomaly in Fig.~\ref{Fig3} is not scaled to real values.

\subsubsection{Spacecraft in Retrograde direction}

In retrograde direction, the new initial conditions are:
\begin{eqnarray}
  v_{apx} &=& V_P + v_{\infty} \cos (\omega \mp \theta) \\
  v_{apy} &=& -v_{\infty} \sin (\omega \mp \theta) \\
  v_{apz} &=& 0
\end{eqnarray}
Therefore
\begin{eqnarray}
  (\mathbf{v} \cdot \mathbf{n}^{'}) &=& v_{apx} \cos (\theta') + v_{apy} \sin(\theta') \\
   &=& V_P \cos (\theta') - v_{\infty} \cos^2 (\theta') + v_{\infty} \sin^2(\theta') \\
    &=&  V_P \cos (\theta') + v_{\infty} \cos 2(\theta')
\end{eqnarray}

In this case, the equation of azimuthal motion is
\begin{equation}\label{eq41}
\frac{d v_{\theta}}{d t}= 2 \omega_{\oplus} v_{\infty} \cos (\theta')\sin I + 2 \omega_{\oplus} \frac{v_{\infty}^2}{c} \cos (\theta') \sin I.
\end{equation}
Now, we may expect that $v_{\infty} \sim v_{\theta}$ and within the range of this assumption the above Eq. becomes
\begin{equation}\label{eq42}
\frac{d v_{\theta}}{d \theta}= K \frac{c}{R_{\oplus}} v_{\infty} \cos (\theta')\sin I  + \frac{Kv_{\infty}^2}{R_{\oplus}} \cos (\theta')^2 \sin I.
\end{equation}
We will make the approach and rewrite the Eq.~\ref{eq42} in the form
\begin{equation}\label{eq43}
\frac{dv_{\theta}}{d \theta'}= \left( \frac{K c v_r}{v} \sin I + \frac{K v_{\infty} v_r}{c} \cos (\theta') \sin I \right) \cos (\theta').
\end{equation}
We make the following approximations:
\begin{eqnarray}
  \frac{K v_{\infty} v_r}{c} &\sim & v_{\theta} \\
  \frac{K c v_r}{v_{\theta}} &\sim & 0.1
\end{eqnarray}
Considering $v_{\infty} \sim 6 \times 10^3$ km$/$s, as reported for the NEAR spacecraft~\cite{Anderson 2008}, we obtain $v_r \sim 7$ km$/$s and $v_{\theta} \sim 1$ km$/$s, values that are in the order of the expected magnitude. The mathematical solution of Eq.~\ref{eq43} is given by
\begin{equation}\label{eq44}
\begin{split}
v_{\theta} = c_1 \exp [c_2 \theta + c_3 \sin (\theta)] (c_4 + \\
c_5 \int_1^{\theta} \exp [-c_2 \xi - c_3 \sin (2 \xi)] \cos \xi d \xi.
\end{split}
\end{equation}

\begin{table}
  \centering
  \caption{Parameters for the flyby velocity anomaly during retrograde or posigrade direction.}\label{Table1}
\begin{tabular}{c c c}
  \hline \hline
\multicolumn{1}{|p{2.4cm}|}{\centering Parameters \\ $c_i$}  & \multicolumn{1}{|p{2.4cm}|}{\centering Retrograde \\ direction} & \multicolumn{1}{|p{2.4cm}|}{\centering Posigrade \\ direction}  \\ [0.25 cm]
\hline \\ [0.25 cm]
  $c_1$   & 1              & 1 \\ [0.25 cm]
  $c_2$   & 0.5            & 0.07 \\[0.25 cm]
  $c_3$   & 0.25           & 0.0233 \\ [0.25 cm]
  $c_4$   & 1              & 1      \\ [0.25 cm]
  $c_5$   & 0.1            & 0.1    \\ [0.25 cm]
  \hline
\end{tabular}
\end{table}

\begin{figure}
  \centering
  \includegraphics[width=3.3 in]{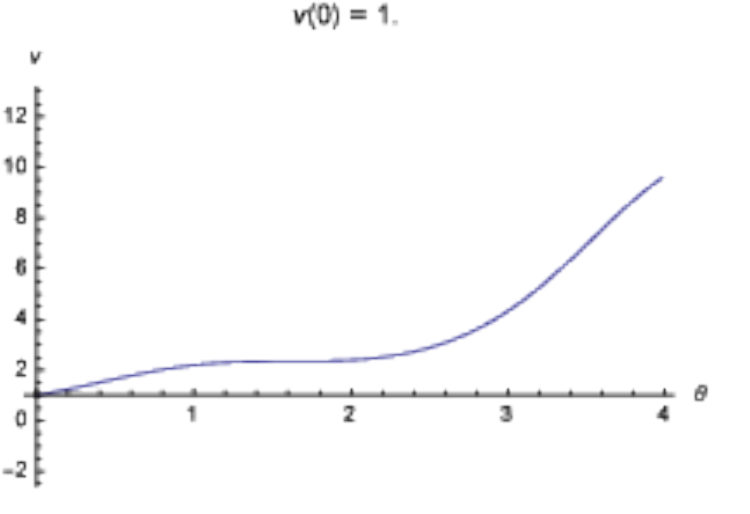}\\
  \caption{Representation of Eq.~\ref{eq44} giving the azimuthal velocity $y=v_{\theta}$ (in m$/$s) vs. $x=\omega \mp \theta$ (in rad units), from $0 \to \pi$, in retrograde motion.}\label{Fig4}
\end{figure}

Fig.~\ref{Fig4} shows a typical solution for the retrograde direction. As before, the magnitude of the velocity anomaly is not scaled to real values.
We may notice that the main difference between the two flyby approaches is the direct dependency on the angle $\theta'=\omega \mp \theta$ in the retrograde case, implying an increase of the magnitude of the velocity anomaly on the outgoing trajectory after the flyby, an asymmetry due to the direct conversion of angular to linear motion intrinsic to the TTC.

\section{Conclusion}

The theory predicts an asymmetry when a spacecraft enters in flyby maneuvering in retrograde or in posigrade direction. The flyby anomaly is the direct outcome of two effects: the relativistic, associate to the time-delay of the gravitational action, and the topological torsion current, which represents a direct conversion of angular to linear motion, which is by itself a new result from the theoretical point of view. The definitive test of the theoretical model should be done with accurate numerical solutions of the proposed set of equations, and comparison with interplanetary spacecraft trajectories data.

\begin{acknowledgments}
The author gratefully acknowledge partial financial support by the Portuguese Funda\c{c}\~{a}o para a Ci\^{e}ncia e Tecnologia under contract ref. SFRH/BSAB/1420/2014.
\end{acknowledgments}


\bibliographystyle{apsrev}
\bibliography{Doc2}
\end{document}